\documentclass[conference]{IEEEtran}
\IEEEoverridecommandlockouts
\usepackage{cite}
\usepackage{amsmath,amssymb,amsfonts}
\usepackage{algorithmic}
\usepackage{graphicx}
\usepackage{textcomp}
\usepackage{xcolor}
\def\BibTeX{{\rm B\kern-.05em{\sc i\kern-.025em b}\kern-.08em
    T\kern-.1667em\lower.7ex\hbox{E}\kern-.125emX}}
\begin{document}
\title{What's behind tight deadlines?\\ Business causes of technical debt
\thanks{ This work is supported by Fundação de Educação Tecnológica e Cultural da Paraíba (Funetec-PB) - contracts ESIG 04/21, Phoebus 04/21, and Snet 04/22.)}
}

\author{

\IEEEauthorblockN{Rodrigo Rebouças de Almeida}
\IEEEauthorblockA{\textit{Federal University of Paraíba} \\
Brazil \\
rodrigor@dcx.ufpb.br}

\and

\IEEEauthorblockN{Christoph Treude}
\IEEEauthorblockA{\textit{The University of Melbourne}\\
Australia \\
christoph.treude@unimelb.edu.au}

\and

\IEEEauthorblockN{Uirá Kulesza}
\IEEEauthorblockA{\textit{Federal University of Rio Grande do Norte} \\
Brazil \\
uira@dimap.ufrn.br}

}

\maketitle

\begin{abstract}
What are the business causes behind tight deadlines? What drives the prioritization of features that pushes quality matters to the back burner? We conducted a survey with 71 experienced practitioners and did a thematic analysis of the open-ended answers to the question: ``Could you give examples of how business may contribute to technical debt?'' Business-related causes were organized into two categories: pure-business and business/IT gap, and they were related to `tight deadlines' and `features over quality', the most frequently cited management reasons for technical debt. We contribute a cause-effect model which relates the various business causes of tight deadlines and the  behavior of prioritizing features over quality aspects.
\end{abstract}


\section{Introduction}\label{sec:introduction}

Technical debt is a concept in software development that refers to the extra development work that is needed to complete a project due to the shortcuts and hacks that were used during the initial development phase~\cite{tom2013exploration}. Technical debt can lead to software evolution problems such as code rot, software bloat, and decreased software quality~\cite{suryanarayana2014refactoring}. It can also make it challenging to add new features or make changes to existing features~\cite{seaman2012using}. Furthermore, technical debt can create a negative feedback loop in which the extra work required to pay off the debt reduces the time available for new development, leading to even more debt~\cite{borowa2021living}.

Business aspects have been identified as a significant force behind the creation of technical debt~\cite{AMPATZOGLOU:2015, KruchtenManagingTechnical2019}. Technical debt has been studied from many perspectives, from code to human behavior aspects. However,  the business dimension lacks deeper exploration~\cite{RIOS2018-tertiary}. Causes of technical debt such as ``tight deadlines'' and ``business pressure'' occupy the top ranks among causes of technical debt~\cite{ErnstBONG15,KruchtenManagingTechnical2019, rios2020}, but what is behind the ``business pressure''? What are the business causes behind tight deadlines? What drives the prioritization of features that pushes quality matters to the back burner?

To understand the impact of how business decisions contribute to technical debt, we conducted a survey with 71 experienced practitioners. Different from previous survey studies~\cite{HolvitieLSHMMBL18, ErnstBONG15}, we focus on the business causes of technical debt. We identify a large number of factors, including market pressure, legal aspects, and gaps between business and IT planning. This survey complements our work on business-driven technical debt management~\cite{icsme2018, techdebt2021} where we found that the business perspective can make a relevant contribution to technical debt management.

\section{Research Method}

The research question we answer in this article is: \textbf{How do business decisions contribute to technical debt?} We analyse answers from 71 participants to the following two questions:

\begin{itemize}
\item Q1: ``To what extent do business decisions lead to the creation of technical debt?" A closed question with five answer options: ``not at all,'' ``to a very small extent,'' ``to some extent,'' ``to a great extent'' and ``to a very great extent.''
\item Q2: ``Could you give examples of how business may contribute to technical debt?", an open-ended question.
\end{itemize}


\textbf{Data Collection}. The survey was primarily publicized on social networks (LinkedIn, Twitter, and Facebook) and via snowballing (i.e., respondents forwarding the survey to other potential respondents). We received 71 anonymous and valid responses.
The responses contained 1644 words (median: 23.5, standard deviation: 23.2).
 The respondents were aware of the TD concept (92\%) and could give concrete examples of technical debt (100\%) after being shown a definition.

\textbf{Data Analysis}. Based on analyzing the survey responses (Q2), we classified the business causes of technical debt into three categories using thematic analysis~\cite{thematic_analysis}. One author coded the answers, and the other two reviewed them. We asked the respondents to give examples of how business may contribute to technical debt. We coded the responses and identified 12 causes divided into three categories: pure business, business/IT gap, and management.

\textbf{Demographics}. The majority of respondents indicated having more than 10 years of experience (63\%), one-quarter of them (25.4\%) have 5-10 years, and only a small amount of people (11.3\%) have 1-4 years of experience. Regarding their professional activities, most of the survey participants have primarily technical responsibilities (66\%), while a quarter (25\%) indicated both technical and business responsibilities, and 9\% pure-business responsibilities. Most of the respondents work for large companies (more than 1,000 employees), in a diverse range of industries, including software (31\%), government (13\%), and finance (8\%). Respondents were located in Brazil (59\%), North America (25\%), and Europe (14\%).

\section{Study Results}

In this section, we present and detail the different categories and kinds of business causes of technical debt found in our study. 

When asked to what extent business decisions lead to the creation of technical debt for the first question (Q1), 96\% of the respondents indicated that business decisions lead to the creation of technical debt (to some extent: 23\%; to a great extent: 51\%; to a very great extent: 23\%) while only 3\% indicated no or low influence.

For Q2, the categories that emerged from the thematic analysis are:
\textbf{pure-business} (i.e., those related to business decision-making and external market forces); causes related to the \textbf{business/IT gap} (i.e., knowledge and planning gap); and \textbf{management}. 

In addition, based on co-occurring mentions in the survey and explicit mentions of cause/effect relationships, we could identify business causes of \textit{tight deadlines} and \textit{feature over quality}, the most cited management causes for technical debt by our respondents. For example, one participant answered that ``[Business deadlines] may press towards [fulfilling requirements as soon as possible] because of [competition].'' From this answer, we could relate the codes \textit{time to market} (from ``business deadlines'') and \textit{rush to deliver to beat competitors}, as a cause of \textit{feature prioritization}.

\subsection{Business causes of technical debt}
\label{section.businesscauses}


The \textbf{management} category (56 codes) included well-known causes of technical debt~\cite{rios2020,prectitionersCauseEffect2021} like \textit{tight deadlines} (19) and the problem of prioritizing \textit{features over quality} (17). Both are aspects that usually are related to short-term benefits. Causes like \textit{bad requirement elicitation} (11) and \textit{changes of development scope within development sprint} (5) were also mentioned. Since the management causes are well discussed~\cite{rios2020}, here we focus on the \textbf{business} and \textbf{business/IT gap} causes behind the two most cited management causes of technical debt: \textbf{tight deadlines} and the prioritization of \textbf{feature over quality}.

\textbf{Tight deadlines} are commonly identified as the top management cause of technical debt~\cite{rios2020,pina21}. With no time left to deliver features, teams must postpone activities in order to meet target releases. Tight deadlines were also the most cited cause of technical debt in our survey. Besides being a cause of technical debt, \textit{tight deadlines} are a consequence of many other problems related to \textbf{pure business} and the \textbf{business/IT gap}. 30\% of the respondents mentioned ``tight deadlines'' as a consequence of other problems.

After \textit{tight deadlines}, \textit{feature over quality} was the second most cited management cause of technical debt. The prioritization of features is often driven by business pressure, like the value perception. For example, ``features create value,'' a business respondent argues that ``A team can invest a week into (i) a new feature that will make 50 million revenue over a year or (ii) can use the same time to make their framework more robust for running regression tests. If the team invests in (ii), that will reflect on the company's quarter results negatively, thus pulling shares down. The team is pressured by finance to put all of its effort in (i).'' 

While the business impact in the context of technical debt is sometimes reduced to tight deadlines~\cite{rios2020}, our analysis reveals a much more complex picture of how external forces and gaps between domains play significant roles in the creation of technical debt. 

There are business causes for technical debt that cannot be avoided, e.g., a business opportunity or a customer's demand, but a subset of business pressures can be reduced if well managed. Our results provide a set of causes behind the two leading management causes for technical debt ``tight deadlines'' and ``feature over quality'' prioritization. \textbf{When we better understand the business causes of technical debt, we can identify problems that could be avoided, thus preventing the creation of technical debt in the first place.}

\begin{figure*}[ht]
\begin{center}
\includegraphics[width=.92\textwidth]{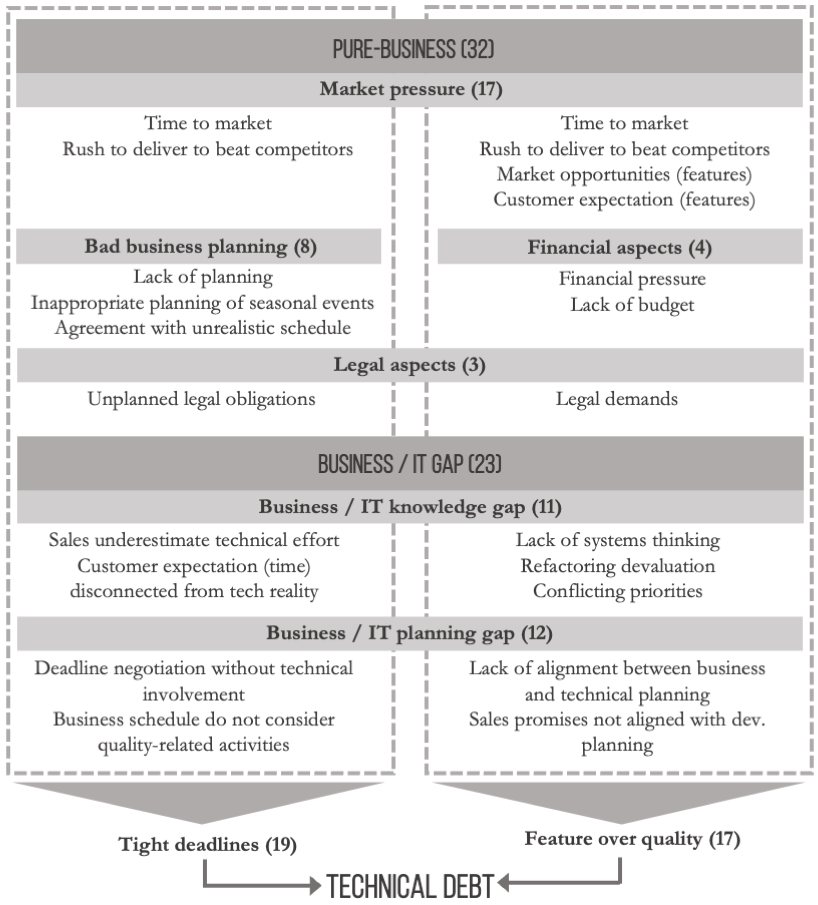}
\caption{Business causes of Technical Debt, with details for the causes of \textit{tight deadlines} and \textit{feature over quality}}
\vspace{-.9cm}
\label{figure-causes-tight} 
\end{center}
\end{figure*}

Figure \ref{figure-causes-tight} presents the code categories of causes of technical debt classified into two main groups: \textbf{Pure-business} (32) and \textbf{Business / IT gap} (23). It also presents the particular causes (the two descending dashed arrows) for \textbf{tight deadlines} (19) and \textbf{feature over quality} (17). For example, \textbf{financial aspects} (4) is a general cause of technical debt, and \textit{financial pressure} plus \textit{lack of budget} were mentioned as financial causes of the prioritization of \textbf{feature over quality}.

In the following, we present the code categories, the number of mentions in the survey, and representative quotes from our respondents.

\subsection{Pure-business}

Pure business aspects are the technical debt causes linked to problems from the business side, like marketing pressure, financial aspects, business planning, legal and political aspects. When the business stakeholders and the client rush for new features and prioritize features over quality, this directly impacts the development schedule. The time to market to beat competitors with new products and features is another point of pressure on development deadlines.

\subsubsection{Market pressure}

The most frequently mentioned pure-business cause for technical debt was market pressure (17 mentions). The market pressure is caused by customers, competitors, opportunities, and time to market.

\textit{Time to market} may cause technical debt by creating forces to release features in a rush to beat the competitors. These forces are causes of \textit{tight deadlines}, and \textit{feature over quality}, like one respondent wrote: ``Releasing features before your competitors may give you business advantages. That could motivate tight deadlines and technical debt.'' 
This ``rush'' to deliver may occur in prototypes planned to be delivered as a production-ready solution, causing \textit{feature over quality} and \textit{tight deadlines}, since the planning does not consider the complete set of features and the quality aspects that should be considered in a production-ready solution. Then, the prototype is shipped as a product with technical debt.

\textit{Market opportunities} and \textit{customer expectations} regarding features to be delivered also play a role in prioritizing \textit{feature over quality} aspects. It is essential to highlight that these are normal and even expected business pressures. However, sometimes new customer demands are delivered to development teams as ``urgent'' without proper prioritization and expectation management. One respondent said that ``Acquisition of new markets, growing the reach of the company/product, which usually leads to a larger cash flow into the company is usually a lot nicer on the eyes of stakeholders than house maintenance, which tends to de-prioritize projects aware of such problems and usually only when shit really hits the fan or things slow down a lot, that's when people review priorities and we end up getting to take time to clean things up.''

Inappropriate management of customer expectations may create unnecessary pressure to deliver a product or service. Arguments like ``the customer wants to see the first version / MVP (Minimum Viable Product) ASAP'' force teams to focus on creating pressure on the delivery of features and postpone work towards quality attributes, thus creating technical debt.

\subsubsection{Bad business planning}

The second most cited pure-business cause for technical debt was \textit{bad business planning} (8). Arbitrary deadlines, bad agenda planning (e.g., \textit{lack of long-term planning for features and projects}), lack of anticipation for recurring business events, frequent ``urgent'' features without previous planning, and hard and difficult-to-negotiate deadlines are some of the mentioned problems that lead to so-called ``business pressure.'' \textit{Inappropriate planning of seasonal events} also creates pressures on deadlines that could be avoided. For example, ``for seasonal marketing and promotional events that happen on the same date, every year, businesses demand new projects and platforms to deal with similar problems over and over again. More than ten years of new `urgent' features coming top-down from business with no planning and no care about tests.''

\subsubsection{Financial aspects}

\textit{Financial pressure} also influences technical debt when there is \textit{lack of budget} to address quality and non-functional requirements (\textit{feature over quality}). As for the ``lack of budgeting for full implementation of solutions,'' sometimes the market demands lots of changes, new features, and prototypes but does not budget for full implementation.

\subsubsection{Legal aspects}

Legal aspects also play a role in creating technical debt since it is an external force that can demand new features under unexpected deadlines. They affect the business schedule and prioritize features (e.g., the new European GDPR privacy law created external demand for systems and services). In our survey, one respondent said that sometimes business planning neglects \textit{legal demands} and schedules, ``often the business demands legal requirements after a legislation term has already expired, creating urgency for the software development.'' Also, the ``lack of alignment between \textit{legal obligations} and the time to demand the changes affected by legal implications'' can put pressure on deadlines and feature prioritization.

\subsection{Business/IT Gap}

Besides the business pressure discussed in the previous section, the gap between business and IT is another cause of technical debt. We categorized the gap into two dimensions: knowledge and planning. The \textit{Business / IT Knowledge gap} encompasses the technical debt causes related to the lack of knowledge from one area about the other. \textit{Business / IT planning gap} is a category of technical debt causes related to deadlines, schedules, planning, and similar concerns.

\subsubsection{Business / IT Knowledge Gap}


Business stakeholders tend to see the development teams as ``black boxes.'' Failing to account for integration concerns, technical impact, and underestimating implementation effort are some of the mentioned causes of technical debt.  Furthermore, the \textit{lack of technical involvement in business decision-making} can lead to ``bad contracts with service providers/partners, leading to integration workarounds.'' Finally, this business/IT gap leads business professionals to ``create product roadmap[s] with little understanding of technology and organizational limitations.''

The knowledge gap between business and IT also contributes to \textit{tight deadlines}. As one respondent stated, ``sales and business analysts underestimate implementation effort.''
 For example, one respondent said that ``salespeople try to sell more than the company can deliver, sales and business analysts underestimate implementation effort or cut down schedules due to client request...''. The \textit{customer expectation} regarding the time when the solution will be delivered is \textit{disconnected from the technical reality}, creating pressure on development deadlines.

The knowledge gap also affects the problem in which features are prioritized over quality aspects. Business stakeholders' \textit{lack of systems thinking} was reported as a cause of feature prioritization, e.g., ``rushing to optimize for one part / one group, resulting in negative side effects to the whole / broader organization.'' Another cause for feature over quality is the \textit{refactoring devaluation}. Some stakeholders do not care about refactoring and other quality aspects, focusing on short-term value delivery. Finally, \textit{conflicting priorities} also contribute to technical debt creation, like ``it has to be done fast, it has to be backward compatible, it has to be future proof.''

\subsubsection{Business / IT Planning Gap}

The \textit{Business / IT planning gap} received 12 mentions. Problems include the lack of alignment between software requirements and technical development, where business participants make commitments that cannot be handled in the expected time. As a result, \textit{sales promises are not aligned with development planning}.


In this category, the problems with \textit{deadline negotiation without technical involvement} and \textit{the business schedule without considering quality-related activities} were identified as causes of \textit{tight deadlines}. - ``Deadlines negotiated on contracts without engineering feedback.'' In addition, the agenda misalignment occurs when the business planning does not consider the technical planning and vice-versa.


Finally, the lack of alignment between business and technical planning may provoke the prioritization of \textit{features over quality} to deliver value. E.g.,
``Business usually affects technical debt when commitments are made without consulting the engineering team. This happens because business is always focused on the value being delivered, while teams focus on delivering value AND reducing the cost of maintaining the product.''


\section{Research agenda}

In this paper, we have scratched the surface of the causes behind tight deadlines and the frequent decision to prioritize features over quality aspects in software development. Business causes are related to different decision-making aspects, considering a wide range of research opportunities, from technical to processes and behavioral aspects. As far as we know, the business causes of technical debt behind the \textit{tight deadlines} and \textit{features over quality} have yet to be explored, and create research opportunities to improve the decision-making and the business and technical process alignment to reduce the creation of technical debt. We suggest research questions in two areas to be explored.

\subsection{Business causes and effects on technical debt}
A first research objective would be to take another step in exploring the cause-effect relationship between business causes in order to better understand the problem and act on its sources. 

\begin{itemize}  

\item What are the business causes and consequences of technical debt?
\item What are the business aspects that influence the pressure over technical teams?
\item How do these business aspects relate to each other?
\end{itemize}

There are numerous sources of business pressure on technical teams, ranging from strategy-related, like marketing expansion or handling competitors, to tactical and operational ones, like human-resources processes, e.g., onboarding and performance evaluation, or HR problems like high team turnover. Previous studies already emphasize the necessity of understanding better and handling the business aspects of technical debt ~\cite{Greville22,holistic,icsme2018}. Identifying and classifying the causes and their consequences are necessary to act on them appropriately.

\subsection{Cognitive bias and technical debt}
Another research objective could be understanding the nature of problems and whether they are related to technical, process, or behavioral issues. 
\begin{itemize}
\item How does cognitive bias in business decision-making influence the creation of technical debt?
\end{itemize}

Many of the business causes presented in Fig. \ref{figure-causes-tight} can also be the consequence of cognitive biases in stakeholders' decision-making.\cite{klara2021} For example, the \textit{agreement with unrealistic schedule} technical debt cause can be a consequence of inappropriate planning, together with anchoring, confirmation, or other cognitive bias~\cite{cognitiveBias}. Research that addresses cognitive biases in decision-making can contribute to reducing the amount of unnecessary pressure and, consequently, reducing technical debt.

\textbf{Working on this research agenda can result} in a set of guidelines to improve the business and IT processes and decision-making to avoid the unnecessary pressure that leads to technical debt. It involves human aspects, reviewing how different stakeholders and their roles can impact day-to-day decision-making. The inappropriate handling of the hierarchy between business and IT stakeholders can force inappropriate assumptions and lead to pressure and lack of negotiation. For example, revising the way many development meetings (such as Scrum ceremonies - sprint planning/review, daily scrum) are conducted and how different participants expose their demands could reduce the amount of pressure.

\section{Discussion and Conclusion}

In this paper, to complement existing work on the numerous technical and operational causes and consequences of technical debt~\cite{RIOS2018-tertiary,rios2020, td-theory, prectitionersCauseEffect2021}, we focus on the business side of what is causing technical debt.

The relationship between business and technical debt presents itself as an intricate web involving many business aspects from different perspectives. Although we organized the causes of technical debt into two big areas and six categories, they are interrelated. Business aspects added to the business/IT gap and management significantly contribute to creating technical debt. 

Market pressures related to customer demands, time to market, and competitors are the leading business causes for technical debt, with the business/IT gap further exacerbating the problem. Tight deadlines were the most cited management cause of technical debt. Going a step further, we uncovered that tight deadlines are caused by a set of pure-business, business/IT gap, and other management causes (see Figure \ref{figure-causes-tight}). The misalignment between decision-making and planning, and the lack of knowledge about technical and business matters are also relevant causes of technical debt. 

It is important to note that many of the presented business-related causes of technical debt cannot be completely avoided. Technical debt provides short-term benefits and incurring debt can be of strategic value, but it must be managed and adequately prioritized to not accumulate over time.

The presented business causes of technical debt, and the most frequent management causes ``tight deadlines'' and ``feature over quality'', can guide decision-making and improve business processes to avoid unnecessary technical debt. 

Practitioners should review the business processes and the decision-making chain and consider paying attention to managing communication and involvement between business and technical teams regarding planning, scope, and effort estimation. Teams should look for ways to prioritize technical debt considering business metrics and perspectives to align business and technical aspects. 

We contribute a cause-effect model (Figure~\ref{figure-causes-tight}), which relates the various business causes of technical debt to each other and explains their impact on technical debt. Practitioners of different roles can use this model to understand the influences on technical debt creation, anticipate issues, and work across business and IT to better manage technical debt.

\subsection{Limitations}
The presented results are based on 71 respondents and cannot be treated as generalizable. To address the sample size limitation, the participants are mostly senior practitioners from diverse companies working in different domains. Surveys are typically subject to sampling bias, namely self-selection bias, which could distort our sample towards the developers who chose to participate. As the majority of respondents have technical responsibilities, the results may skew towards a technical perspective. All codes and categorizations were reviewed by at least one author not involved in the coding, and any conflicts were solved by a second author. There is no distinction between different types of technical debt regarding the presented causes. The causes were declared independent of the debt type.



\bibliographystyle{IEEEtran}
\bibliography{body/techdebt}

\begin{thebibliography}{10}
\providecommand{\url}[1]{#1}
\csname url@samestyle\endcsname
\providecommand{\newblock}{\relax}
\providecommand{\bibinfo}[2]{#2}
\providecommand{\BIBentrySTDinterwordspacing}{\spaceskip=0pt\relax}
\providecommand{\BIBentryALTinterwordstretchfactor}{4}
\providecommand{\BIBentryALTinterwordspacing}{\spaceskip=\fontdimen2\font plus
\BIBentryALTinterwordstretchfactor\fontdimen3\font minus
  \fontdimen4\font\relax}
\providecommand{\BIBforeignlanguage}[2]{{%
\expandafter\ifx\csname l@#1\endcsname\relax
\typeout{** WARNING: IEEEtran.bst: No hyphenation pattern has been}%
\typeout{** loaded for the language `#1'. Using the pattern for}%
\typeout{** the default language instead.}%
\else
\language=\csname l@#1\endcsname
\fi
#2}}
\providecommand{\BIBdecl}{\relax}
\BIBdecl

\bibitem{tom2013exploration}
E.~Tom, A.~Aurum, and R.~Vidgen, ``An exploration of technical debt,''
  \emph{Journal of Systems and Software}, vol.~86, no.~6, pp. 1498--1516, 2013.

\bibitem{suryanarayana2014refactoring}
G.~Suryanarayana, G.~Samarthyam, and T.~Sharma, \emph{Refactoring for software
  design smells: managing technical debt}.\hskip 1em plus 0.5em minus
  0.4em\relax Morgan Kaufmann, 2014.

\bibitem{seaman2012using}
C.~Seaman, Y.~Guo, N.~Zazworka, F.~Shull, C.~Izurieta, Y.~Cai, and
  A.~Vetr{\`o}, ``Using technical debt data in decision making: Potential
  decision approaches,'' in \emph{2012 Third International Workshop on Managing
  Technical Debt (MTD)}.\hskip 1em plus 0.5em minus 0.4em\relax IEEE, 2012, pp.
  45--48.

\bibitem{borowa2021living}
K.~Borowa, A.~Zalewski, and A.~Saczko, ``Living with technical debt—a
  perspective from the video game industry,'' \emph{IEEE Software}, vol.~38,
  no.~06, pp. 65--70, 2021.

\bibitem{AMPATZOGLOU:2015}
A.~Ampatzoglou, A.~Ampatzoglou, A.~Chatzigeorgiou, and P.~Avgeriou, ``The
  financial aspect of managing technical debt: A systematic literature
  review,'' \emph{Information and Software Technology}, vol.~64, pp. 52--73,
  2015.

\bibitem{KruchtenManagingTechnical2019}
P.~Kruchten, R.~Nord, and I.~Ozkaya, \emph{Managing Technical Debt: Reducing
  Friction in Software Development}.\hskip 1em plus 0.5em minus 0.4em\relax
  Software Engineering Institute, Carnegie Mellon University, 2019.

\bibitem{RIOS2018-tertiary}
N.~Rios, M.~G. de~Mendon{\c c}a~Neto, and R.~O. Sp{\'\i}nola, ``A tertiary
  study on technical debt: Types, management strategies, research trends, and
  base information for practitioners,'' \emph{Information and Software
  Technology}, vol. 102, pp. 117--145, 2018.

\bibitem{ErnstBONG15}
N.~A. Ernst, S.~Bellomo, I.~Ozkaya, R.~L. Nord, and I.~Gorton, ``Measure it?
  manage it? ignore it? software practitioners and technical debt,'' in
  \emph{Proceedings of the 2015 10th Joint Meeting on Foundations of Software
  Engineering, {ESEC/FSE} 2015, Bergamo, Italy, August 30 - September 4, 2015},
  E.~D. Nitto, M.~Harman, and P.~Heymans, Eds.\hskip 1em plus 0.5em minus
  0.4em\relax {ACM}, 2015, pp. 50--60.

\bibitem{rios2020}
N.~Rios, R.~O. Sp{\'\i}nola, M.~Mendon{\c c}a, and C.~Seaman, ``The
  practitioners' point of view on the concept of technical debt and its causes
  and consequences: a design for a global family of industrial surveys and its
  first results from brazil,'' \emph{Empirical Software Engineering}, vol.~25,
  09 2020.

\bibitem{HolvitieLSHMMBL18}
J.~Holvitie, S.~A. Licorish, R.~O. Sp{\'{\i}}nola, S.~Hyrynsalmi, S.~G.
  MacDonell, T.~S. Mendes, J.~Buchan, and V.~Lepp{\"{a}}nen, ``Technical debt
  and agile software development practices and processes: An industry
  practitioner survey,'' \emph{Inf. Softw. Technol.}, vol.~96, pp. 141--160,
  2018.

\bibitem{icsme2018}
R.~{Rebou{\c c}as de Almeida}, U.~{Kulesza}, C.~{Treude}, D.~{Cavalcanti
  Feitosa}, and A.~{Higino Guedes Lima}, ``Aligning technical debt
  prioritization with business objectives: A multiple-case study,'' in
  \emph{Proc.~of the Int'l.~Conf.~on Software Maintenance and Evolution -
  ICSME}, 2018, pp. 655--664.

\bibitem{techdebt2021}
R.~{Rebou{\c c}as de Almeida}, C.~{Treude}, and U.~{Kulesza}, ``Business-driven
  technical debt prioritization: An industrial case study,'' in \emph{Proc.~of
  the Int'l.~Conf.~on International Conference on Technical Debt -
  TechDebt2021}, 2021.

\bibitem{thematic_analysis}
V.~Braun and V.~Clarke, ``Using thematic analysis in psychology,''
  \emph{Qualitative Research in Psychology}, vol.~3, no.~2, pp. 77--101, 2006.

\bibitem{prectitionersCauseEffect2021}
S.~Freire, N.~Rios, B.~Perez, C.~Castellanos, D.~Correal, R.~Rama{\v{c}},
  V.~Mandi{\'{c}}, N.~Tau{\v{s}}an, G.~L{\'{o}}pez, A.~Pacheco, D.~Falessi,
  M.~Mendon{\c{c}}a, C.~Izurieta, C.~Seaman, and R.~Spinola, ``{How Experience
  Impacts Practitioners' Perception of Causes and Effects of Technical Debt},''
  in \emph{14th International Conference on Cooperative and Human Aspects of
  Software Engineering (CHASE 2021)}, 2021.

\bibitem{pina21}
D.~Pina, A.~Goldman, and G.~Tonin, ``Technical debt prioritization: Taxonomy,
  methods results, and practical characteristics,'' in \emph{2021 47th
  Euromicro Conference on Software Engineering and Advanced Applications
  (SEAA)}, 2021, pp. 206--213.

\bibitem{Greville22}
M.~Greville, P.~O'Raghallaigh, and S.~McCarthy, ``A triple bottom-line typology
  of technical debt: Supporting decision- making in cross-functional teams,''
  in \emph{55th Hawaii International Conference on System Sciences, {HICSS}
  2022, Virtual Event / Maui, Hawaii, USA, January 4-7, 2022}.\hskip 1em plus
  0.5em minus 0.4em\relax ScholarSpace, 2022, pp. 1--10.

\bibitem{holistic}
S.~Malakuti and J.~Heuschkel, ``The need for holistic technical debt management
  across the value stream: Lessons learnt and open challenges,'' in \emph{2021
  IEEE/ACM International Conference on Technical Debt (TechDebt)}, 2021, pp.
  109--113.

\bibitem{klara2021}
K.~Borowa, A.~Zalewski, and S.~Kijas, ``The influence of cognitive biases on
  architectural technical debt,'' in \emph{2021 IEEE 18th International
  Conference on Software Architecture (ICSA)}, 2021, pp. 115--125.

\bibitem{cognitiveBias}
A.~Caputo, ``A literature review of cognitive biases in negotiation
  processes,'' \emph{International Journal of Conflict Management}, vol.~24,
  no.~4, pp. 374--398, 2022/11/18 2013.

\bibitem{td-theory}
R.~Verdecchia, P.~Kruchten, and P.~Lago, ``Architectural technical debt: A
  grounded theory,'' in \emph{Software Architecture}, A.~Jansen, I.~Malavolta,
  H.~Muccini, I.~Ozkaya, and O.~Zimmermann, Eds.\hskip 1em plus 0.5em minus
  0.4em\relax Cham: Springer International Publishing, 2020, pp. 202--219.

\end{thebibliography}

\end{document}